%% file: hicss.tex
\documentclass[10pt]{article}
\input{hicss-packages.tex}

\usepackage{acro}

\usepackage[colorlinks=true, linkcolor=blue, citecolor=blue, urlcolor=blue]{hyperref}

\usepackage{ifpdf}
\usepackage{amsmath,amssymb,amsfonts}
\usepackage{algorithmic}
\usepackage{graphicx,color}
\usepackage{textcomp}
\usepackage{hyperref}
\usepackage[ruled,vlined,linesnumbered]{algorithm2e}
\usepackage{multirow}
\usepackage{subcaption}
\usepackage{placeins}
\usepackage{float}
\usepackage{multicol}
\usepackage{pifont}
\usepackage{textcomp}
\usepackage{etoolbox}

\usepackage{array}
\AtBeginBibliography{\fontsize{9pt}{11pt}\selectfont}
\usepackage{caption}
\usepackage{subcaption}

\title{Emission-Aware Operation of Electrical Energy Storage Systems}

\author{Haotian Yao \\
University of Calgary \\
{\underline{ haotian.yao@ucalgary.ca}} \\ \\
Mostafa Farrokhabadi \\
University of Calgary \\
{\underline{ mostafa.farrokhabadi@ucalgary.ca} } \\ \And
Vahid Hakimian \\
University of Calgary \\
{\underline{ vahid.hakimian@ucalgary.ca} } \\ \\
Hamidreza Zareipour\\
University of Calgary \\
{\underline{ hzareipo@ucalgary.ca} } \\ }

\date{}

\DeclareAcronym{DERs}{
    short = DERs,
    long = Distributed Energy Resources,
    first-style = long-short
}

\DeclareAcronym{EDS}{
    short = EDS,
    long = electrical distribution system,
    first-style = long-short
}

\DeclareAcronym{RES}{
    short = RES,
    long = Renewable Energy Sources,
    first-style = long-short
}

\DeclareAcronym{EVs}{
    short = EVs,
    long = Electric Vehicles,
    first-style = long-short
}

\DeclareAcronym{PV}{
    short = PV,
    long = Photovoltaics,
    first-style = long-short
}

\DeclareAcronym{ESS}{
    short = ESS,
    long = energy storage systems,
    first-style = long-short
}

\DeclareAcronym{ESP}{
    short = ESP,
    long = Energy Storage Providers,
    first-style = long-short
}

\DeclareAcronym{TEM}{
    short = TEM,
    long = transactive energy market,
    first-style = long-short
}

\DeclareAcronym{TCCEEM}{
    short = TCCEEM,
    long = Transactive Coupled Carbon Emission and Energy Market,
    first-style = long-short
}

\DeclareAcronym{P2P}{
    short = P2P,
    long = Peer-to-Peer,
    first-style = long-short
}

\DeclareAcronym{LME}{
    short = LME,
    long = Locational Marginal Emission,
    first-style = long-short
}

\DeclareAcronym{DLME}{
    short = DLME,
    long = Distribution Locational Marginal Emission,
    first-style = long-short
}

\DeclareAcronym{LMP}{
    short = LMP,
    long = Locational Marginal Price,
    first-style = long-short
}

\DeclareAcronym{DLMP}{
    short = DLMP,
    long = Distribution Locational Marginal Price,
    first-style = long-short
}

\DeclareAcronym{CDLMP}{
    short = CDLMP,
    long = Carbon-Aware Distribution Locational Marginal Pricing,
    first-style = long-short
}

\DeclareAcronym{VPPs}{
    short = VPPs,
    long = Virtual Power Plants,
    first-style = long-short
}

\DeclareAcronym{DSO}{
    short = DSO,
    long = Distributed System Operator,
    first-style = long-short
}

\DeclareAcronym{ISO}{
    short = ISO,
    long = Independent System Operator,
    first-style = long-short
}

\DeclareAcronym{OPF}{
    short = OPF,
    long = Optimal Power Flow,
    first-style = long-short
}

\DeclareAcronym{ACOPF}{
    short = ACOPF,
    long = Alternating Current Optimal Power Flow,
    first-style = long-short
}

\DeclareAcronym{TIER}{
    short = TIER,
    long = Technology Innovation and Emissions Reduction,
    first-style = long-short
}

\DeclareAcronym{FPPs}{
    short = FPPs,
    long = Federated Power Plants,
    first-style = long-short
}

\DeclareAcronym{SR-MEF}{
    short = SR-MEF,
    long = Short-Run Marginal Emissions Factors,
    first-style = long-short
}

\DeclareAcronym{AEF}{
    short = AEF,
    long = Average Emissions Factors,
    first-style = long-short
}

\DeclareAcronym{CHP}{
    short = CHP,
    long = Combined Heat and Power,
    first-style = long-short
}

\DeclareAcronym{SOCP}{
    short = SOCP,
    long = Second-Order Cone Programming,
    first-style = long-short
}

\DeclareAcronym{V2G}{
    short = V2G,
    long = Vehicle-to-Grid,
    first-style = long-short
}

\DeclareAcronym{GenCos}{
    short = GenCos,
    long = Generation Companies,
    first-style = long-short
}

\DeclareAcronym{CEA}{
    short = CEA,
    long = Carbon Emission Allowances,
    first-style = long-short
}

\DeclareAcronym{EPCs}{
    short = EPCs,
    long = emission performance credits,
    first-style = long-short
}

\DeclareAcronym{DGs}{
    short = DGs,
    long = Distributed Generations,
    first-style = long-short
}

\DeclareAcronym{DSM}{
    short = DSM,
    long = Demand-Side Management,
    first-style = long-short
}

\DeclareAcronym{AMS}{
    short = AMS,
    long = Active Management Schemes,
    first-style = long-short
}

\DeclareAcronym{EMS}{
    short = EMS,
    long = energy management system,
    first-style = long-short
}

\DeclareAcronym{MILP}{
    short = MILP,
    long = Mixed-Integer Linear Programming,
    first-style = long-short
}

\DeclareAcronym{PSO}{
    short = PSO,
    long = Particle Swarm Optimization,
    first-style = long-short
}

\DeclareAcronym{BTM}{
    short = BTM,
    long = Behind-The-Meter,
    first-style = long-short
}

\DeclareAcronym{ADNs}{
    short = ADNs,
    long = Active Distribution Networks,
    first-style = long-short
}

\DeclareAcronym{ADMM}{
    short = ADMM,
    long = Alternating Direction Method of Multipliers,
    first-style = long-short
}

\DeclareAcronym{MEI}{
    short = MEI,
    long = marginal emission intensity,
    first-style = long-short
}

\DeclareAcronym{CEF}{
    short = CEF,
    long = Carbon Emission Flow,
    first-style = long-short
}

\begin{document}
\maketitle
\begin{abstract}

Since the beginning of this century, there has been a growing body of research and developments supporting the participation of energy storage systems (ESS) in the emission reduction mandates. However, regardless of these efforts and despite the need for an accelerated energy transition, we have yet to see a practical framework for operational carbon accounting and credit trading for energy storage systems. In this context, this paper proposes an emission performance credits (EPCs) framework that allows ESS, down to the prosumer level, to participate in the carbon market. Thus, a mechanism is proposed, for the first time, to calculate the grid's real-time marginal emission intensity (MEI). The MEI is then used to optimize the cumulative operational emission of ESS through carbon-aware dispatch. Consequently, the framework tracks the operational emissions and converts them into EPCs, which are then sold to regulated entities under compliance programs. Simulation results support the potential of ESS, regardless of their size, to participate in the broader carbon mitigation objectives.

\end{abstract}

\subsubsection*{Keywords:}

Energy storage system, carbon accounting, credit trading, marginal emission intensity, optimal dispatch.

\section{Introduction}

Strong global growth has been observed across all applications of \ac{ESS} within the past five years, with a total of 120 GW of new capacity added worldwide (\cite{IEA_BESS}). Utility-scale \ac{ESS}, typically installed at transmission or distribution nodes, provides services such as frequency regulation, peak shaving, load shifting, and \ac{RES} integration (\cite{9740444, 9166729, 8003298}). These are often co-optimized with generation and transmission infrastructure in wholesale electricity markets towards multiple objectives, including enhanced efficiency, economics, and security (\cite{9089020}). Smaller-scale \ac{ESS}, deployed at the distribution level, support local operational objectives, e.g., voltage regulation, and may also participate in wider grid ancillary services such as frequency regulation (\cite{9557813}). At the behind-the-meter level, residential \ac{ESS} are used for energy arbitrage and self-consumption maximization (SCM) of \ac{RES} generation (\cite{7999305, azuatalam2018techno}). A common gap among the above-mentioned studies, and many of other ESS applications literature, is the absence of emission considerations (\cite{8610327}).

A survey of prior art on the intersection of ESS and emission reduction or carbon accounting returns very few works. Within this limited body of work, the majority focus on co-optimizing integrated systems of \ac{ESS} and \ac{RES}, e.g., see \cite{lin2016emissions, hittinger2015bulk, babacan2018unintended, feng2022bi, 8844848, zafirakis2015embodied}. In such frameworks, emission reductions are primarily attributed to reduced curtailment of \ac{RES}, rather than the direct emissions impact of \ac{ESS} operation itself (\cite{du2024real}). While valuable, these approaches do not fully capture the standalone emission reduction potential of \ac{ESS} when strategically dispatched in response to grid carbon intensity and carbon pricing signals. 

A few works have considered the standalone impact of ESS operational emissions (\cite{colbert2021greenhouse,10433421,Gu2023790}). For instance, \cite{10433421} introduces the concept of a cleanness value to quantify the emission reduction enabled by spatiotemporal \ac{ESS} dispatch, while \cite{Gu2023790} proposes a chronological model to reflect the influence of \ac{ESS} on the evolution of \ac{CEF} over time. However, these studies rely on average emission intensity metrics and thus fail to capture real-time marginal emissions. Moreover, they also focus primarily on the co-optimization of \ac{ESS} and \ac{RES} within distribution networks rather than enabling direct \ac{ESS} participation in carbon markets.

Inadequate mechanisms for ESS participation in the carbon market pose a significant lost opportunity. This gap exists due to \ac{ESS}' negligible operational emissions. Hence, they fall outside the scope of traditional carbon market regulations, primarily designed to target direct emitters such as fossil-fueled power plants and industrial facilities (\cite{government2019technology, EUETS}). However, carbon pricing is well-suited for \ac{ESS} due to their operational flexibility and responsiveness to market signals. In fact, market-based carbon pricing mechanisms can significantly benefit from participants' flexibility (\cite{piperagkas2011stochastic}), as conventional participants are generally operationally rigid (\cite{9770947}).   

Participation in market-based carbon pricing mechanisms, such as a cap-and-trade system, can simultaneously reduce the carbon footprint of ESS while enhancing their operational economics. Thus, we propose that \ac{ESS} participation in these markets can significantly influence system-wide emissions by charging during low-emission periods and discharging otherwise. In this context, the lack of standardized methods to account for the indirect yet significant emissions associated with \ac{ESS} remains a key barrier to its participation in compliance-based carbon markets, and is a key contribution of this work. Existing approaches for carbon emissions accounting generally fall into two categories: (1) average emission intensity, such as the \ac{CEF} approach, which estimate the average emissions associated with electricity consumption (\cite{7021901}); and (2) \ac{MEI}, which capture the change in system-wide emissions resulting from incremental increase in electricity demand (\cite{he2024locational}). While average-based methods are relatively simple to implement, they do not capture the dynamic, time-sensitive emissions impact of dispatch decisions. In contrast, MEI provides a more accurate and granular signal, making it particularly suitable for guiding the real-time operation of flexible assets like \ac{ESS}.

Despite the intuitive alignment of market-based carbon pricing and ESS operational capabilities, carbon accounting specifically tailored to \ac{ESS} remains limited due to the aforementioned gaps (\cite{Jiang20234724}). To address these, this paper proposes, first, a framework for grid MEI calculation that is aligned with \ac{ESS} operation and takes into account the marginal generation rather than the demand-side impact; second, it provides an analysis of the \ac{ESS} operational emission impact and the associated economic and technical opportunities using real-world data. The analysis is based on a conventional arbitrage mechanism, using real-time electricity prices and the proposed calculated grid \ac{MEI} as decision signals. This allows the environmental impact of \ac{ESS} operation to be accurately captured and linked to tradable \ac{EPCs} in the carbon market. Thus, the main contributions of this paper are as follows:

\begin{itemize}
    \item A practical and scalable method to calculate real-time \ac{MEI} in \ac{ESS} operation by identifying the responsiveness of load-following generators to marginal residual demand changes.

    \item A numerical analysis of the opportunities of ESS direct participation in carbon markets leveraging the proposed \ac{MEI} calculation framework using real-world data.

\end{itemize}

The rest of the paper is organized as follows: Section 2 formulates the carbon market and optimization framework; it also provides the calculation process of \ac{MEI}. Section 3 presents data and results. Conclusions and future works are provided in Section 4.

\section{Carbon-Aware ESS Operation}

The proposed framework enables the carbon-aware \ac{ESS} operation and is agnostic to the ESS size or grid-connection level. Thus, the ESS can participate in carbon markets through the generation and trading of \ac{EPCs}\footnote{Carbon markets subject regulated facilities to mandatory regulations to meet prescribed emissions benchmarks. Facilities that reduce emissions beyond their regulatory obligations can generate \ac{EPCs}, which represent verified avoided carbon emissions resulting from emission-shifting operation. Once quantified, \ac{EPCs} are treated as a tradable asset. They can be sold to regulated entities to offset their excess emissions and meet regulatory targets. The \ac{EPCs} are valued based on prevailing carbon prices, establishing a monetary incentive for market participants to reduce emissions.}. Without the loss of generality, a generic \ac{ESS} dispatch model is considered that abstracts away network-specific constraints in order to isolate the emission reduction potential of carbon-aware scheduling. This abstraction applies equally to grid-scale storage or aggregated smaller-scale systems, and assumes sufficient controllability and market access for participation in compliance carbon markets. The authors acknowledge that such simplifications may undermine the practicality of the proposed method and thus defer such consideration to future works.


\subsection{Marginal Emission Intensity}
%
%
%
%
%

Without loss of generality, this study uses data from Ontario, Canada to develop and verify the proposed approach. Accroding to the Ontario Independent Electricity System Operator (IESO), transmission-level generation is broadly classified into three categories: (1) baseload generation, including nuclear and run-of-the-river hydro, which operates continuously and is largely unresponsive to short-term demand changes; (2) variable generation, such as wind and solar, whose output is weather-dependent but can be curtailed if needed; and (3) intermediate and peaking generation, primarily consisting of natural gas and dispatchable hydro, which are capable of adjusting their output in real time to balance supply and demand. Considering the grid control logic, the latter group, as well as, net imports from neighboring jurisdictions, are identified as marginal units; these units are dispatched to respond to short-term changes in demand.  

In order to isolate the influence of renewable and baseload generation, the residual demand is used to distinguish changes in marginal sources that are indeed driven by demand fluctuations, rather than by the non-dispatchable energy resources. Residual demand is defined as the total Ontario demand minus the output from less flexible generation sources, including nuclear, baseload hydro, wind, solar, and bioenergy (\cite{MEI2020}). The hourly residual demand is calculated as:

\begin{equation}
\begin{aligned}
P _{t}^{\text{R,D}} = P _{t}^{\text{T,D}} - \sum_{m \in \mathcal{M}} P _{m,t} \\
\forall t \in \mathcal{T}
\end{aligned}
\label{Res_Demand}
\end{equation}

\noindent where $\mathcal{T}$ denotes the set of time intervals over the dispatch horizon, and $\mathcal{M}$ is the set of non-dispatchable resources. $P_{t}^{\text{T,D}}$ and $P_{t}^{\text{R,D}}$ are the total Ontario demand and the residual demand at time $t$, respectively. $P _{m,t}$ denotes the hourly output from baseload generation, including nuclear, wind, solar, and bioenergy \footnote{Baseload hydro generation is not excluded from total demand in (\ref{Res_Demand}), as the Ontario dataset does not explicitly distinguish between baseload and dispatchable hydro. The implications of using total hydro generation, rather than isolating peaking hydro, on the residual demand relationship are examined in a subsequent section.}.

\begin{figure}[t!]
    \centering

    \begin{subfigure}[b]{0.45\textwidth}
        \includegraphics[width=\linewidth]{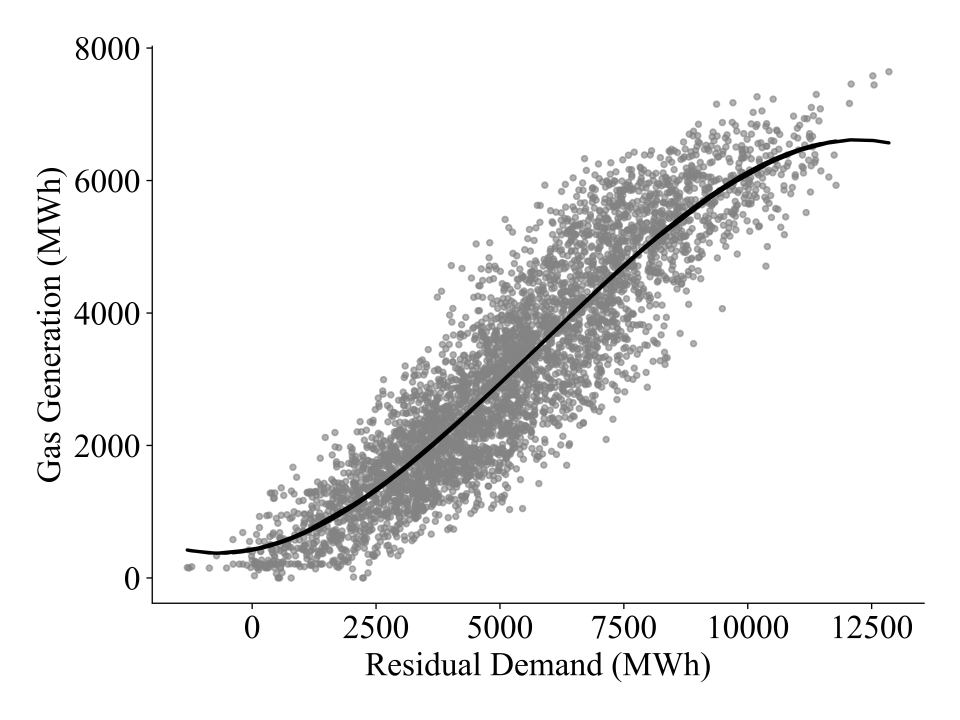}
        \caption{Gas Generation}
        \label{fig:Gas}
    \end{subfigure}
    \hfill
    \begin{subfigure}[b]{0.45\textwidth}
        \includegraphics[width=\linewidth]{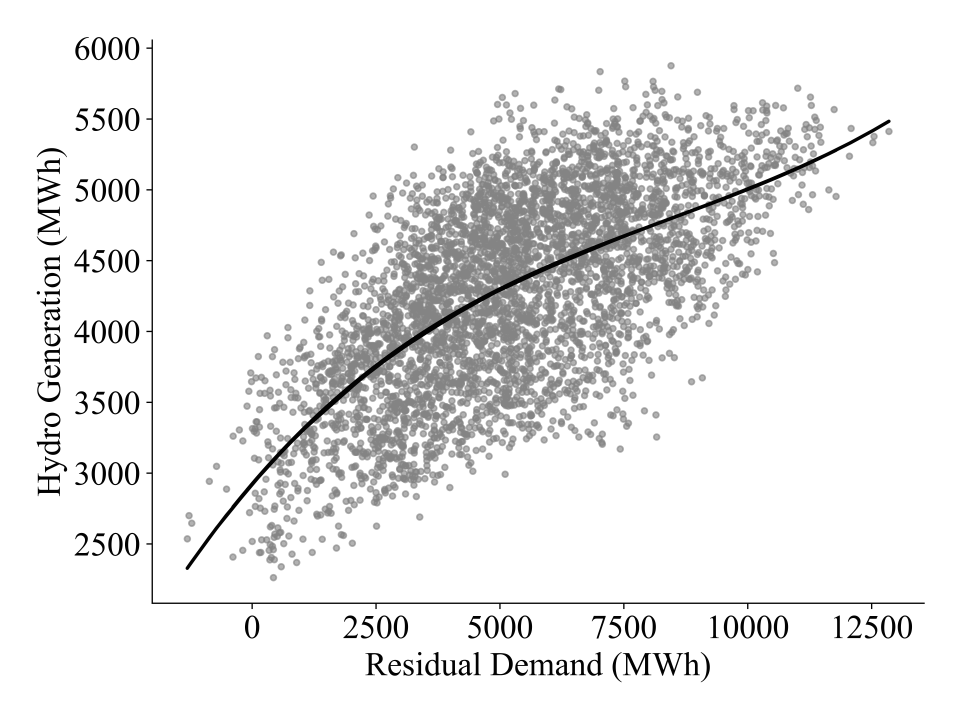}
        \caption{Hydro Generation}
        \label{fig:Hydro}
    \end{subfigure}

    \begin{subfigure}[b]{0.45\textwidth}
        \includegraphics[width=\linewidth]{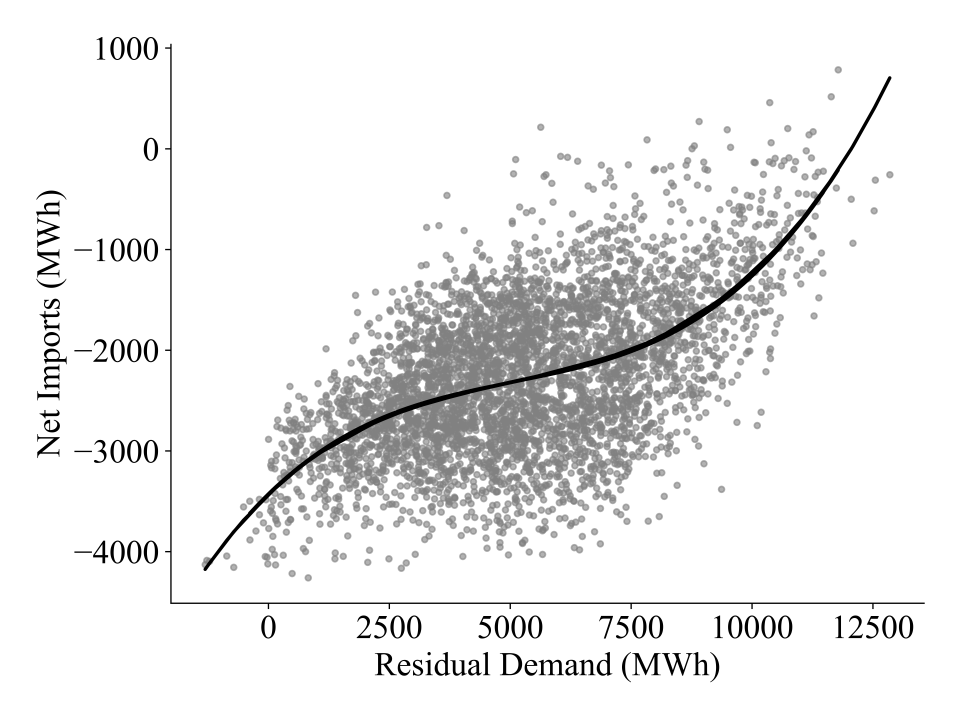}
        \caption{Net Imports}
        \label{fig:Import}
    \end{subfigure}

    \caption{Marginal units output vs.\ residual demand.}
    \label{fig:Marginal_Res}
\end{figure}

To examine how marginal resources respond to variations in the residual demand, hourly data from October 2024 to April 2025 is obtained from the IESO public data directory (\cite{IESOReports})\footnote{IESO pubic data directory includes data up to three months prior to the present. The authors have been downloading this data continually for the past eight months. At the time of submission, the authors have gathered around six months of data, and thus future work is expected to be based on datasets with a longer duration.}. Details of this data is further discussed in Section 3. The data is used to model marginal units' response using cubic regression, as illustrated in Figure~\ref{fig:Marginal_Res}. The accuracy of each model is assessed using the coefficient of determination $R^2$.

As shown in Figure \ref{fig:Gas}, gas generation demonstrates a strong correlation with the residual demand, characterized by a sigmoid-shaped cubic regression curve and a high $R^2$ value of 0.824. At low residual demand levels, gas generation remains near its practical minimum, indicating limited marginal activity when baseload generation and renewable resources are abundant. As residual demand rises, gas generation also increases, reflecting its primary role as Ontario’s marginal supply resource. At high residual demand levels, the curve begins to plateau, suggesting that available gas capacity is scarce, and additional increases in demand must be met through other means.

Hydro generation, in contrast, follows a smoother and more gradual upward trajectory across the full range of residual demand, as illustrated in Figure \ref{fig:Hydro}. The correlation, while positive, is more subtle than gas generation. Notably, Hydro generation begins from a baseline of approximately 2,500 MW, reflecting the contribution of Ontario’s baseload hydro. The $R^2$ value of 0.437 indicates a moderate correlation, consistent with hydro’s more constrained role in marginal balancing. This reflects the dual nature of hydro: a significant portion operates as baseload, while a smaller dispatchable component serves a marginal function, particularly during low and moderate demand conditions. However, its contribution is typically bounded by hydrological constraints and system regulation policies.

On the other hand, net imports, illustrated in Figure \ref{fig:Import}, exhibit the least correlated variable of the three sources, signified by a lower $R^2$ value of 0.314. Despite this, a modest increasing trend can be observed. At high demand levels, net imports become distinctly positive and more strongly correlated with residual demand, suggesting that external supply becomes increasingly important when internal dispatchable resources approach full utilization. These observations align with the intuition that the import behavior is influenced more by external market conditions, thus limiting its effectiveness as a marginal resource.

Overall, the three fitted curves collectively describe Ontario’s marginal supply strategy. Natural gas provides the most elastic and dominant response to load fluctuations, hydro offers a constrained yet consistent contribution, and imports act as a flexible, though externally influenced, marginal resource.

To model the behavior of marginal resources,
the fitted curves are approximated piece-wise linearly into 15 segments. The residual demand less than -1,000 MWh forms the first segment, and the segmentation proceeds thereafter in increments of 1,000 MWh up to values greater than or equal to 12,000 MWh. For each segment $s$, Table~\ref{tab:MEI} summarizes the slopes of the linear regression for each resource type, representing the rate of change in generation or imports per 1 MWh change in the residual demand. In this Table, $\lambda_{n,s}$ is the marginal supply share of resource $n$ in segment $s$, and the \ac{MEI} $m_s$ for each segment $s$ is defined as follows:

\begin{equation}
m _{s} = \sum_{n \in \mathcal{N}} \lambda_{n,s} e_n
\label{eq:MEI_segment}
\end{equation}

\noindent where $e_n$ is the emission factor of resource $n$. In this study, the emission factors for gas and imports are assumed to be 0.37 tCO$_2$e/MWh and 0.44 tCO$_2$e/MWh, respectively, while the emission factor for hydro is set to zero (\cite{ElectricityMaps,MEI2020}). Imports are considered to contribute to the \ac{MEI} only when net imports are positive, primarily during periods of very high residual demand, as illustrated in Figure~\ref{fig:Import}. Consequently, the emission factor for imports, as reflected in Table \ref{tab:MEI}, is incorporated into the \ac{MEI} calculation only for segments 14 and 15. For the majority of residual demand levels, corresponding to segments 1 through 13 where net imports are negative, the \ac{MEI} values are determined solely by gas generation.

As it can be seen in Table \ref{tab:MEI}, the fifth column reports the total marginal supply response, calculated as the sum of the marginal slopes of all three resources. This value remains approximately equal to one across segments, verifying the proposed approach and that nearly the entire marginal demand is met by the combined contributions of gas, hydro, and imports. Thus, for each hour $t$, the \ac{MEI} is calculated as follows:

\begin{equation}
\rho _{t}^{\text{G}} = m _{s} \quad \text{where} \quad P _{t}^{\text{R,D}} \in \text{Segment } s
\label{eq:MEI_time}
\end{equation}

\begin{table}[thb]
\centering
\caption{Ontario marginal supply shares and estimated \ac{MEI} by residual demand segment.}
\begin{tabular}{l|rrrr|>{\centering\arraybackslash}p{1.5cm}}
\hline
\multirow{2}{*}{$s$} & \multicolumn{4}{c|}{Supply share $\lambda_{n,s}$ (MWh)} & {\ac{MEI} $m _{s}$} \\ \cline{2-5}
                   & Gas & Hydro & Import & Total & tCO$_2$e/MWh \\
\hline
1  & -0.144 & 0.500 & 0.677 & 1.033 & -0.053 \\
2  &  0.054 & 0.437 & 0.531 & 1.023 &  0.020 \\
3  &  0.236 & 0.377 & 0.400 & 1.013 &  0.087 \\
4  &  0.407 & 0.318 & 0.280 & 1.004 &  0.151 \\
5  &  0.540 & 0.267 & 0.190 & 0.997 &  0.200 \\
6  &  0.637 & 0.224 & 0.131 & 0.992 &  0.236 \\
7  &  0.698 & 0.189 & 0.101 & 0.988 &  0.258 \\
8  &  0.726 & 0.163 & 0.098 & 0.987 &  0.269 \\
9  &  0.718 & 0.144 & 0.125 & 0.987 &  0.266 \\
10 &  0.677 & 0.133 & 0.179 & 0.989 &  0.250 \\
11 &  0.599 & 0.129 & 0.265 & 0.993 &  0.222 \\
12 &  0.485 & 0.134 & 0.379 & 0.998 &  0.179 \\
13 &  0.344 & 0.146 & 0.515 & 1.005 &  0.127 \\
14 &  0.178 & 0.164 & 0.671 & 1.013 &  0.361 \\
15 & -0.049 & 0.194 & 0.880 & 1.025 &  0.369 \\
\hline
\end{tabular}
\label{tab:MEI}
\end{table}

\subsection{Optimization Framework}

The \ac{ESS} carbon-aware dispatch structure is illustrated in Figure~\ref{fig: Framework}. In this figure, $\lambda_{t}^{\text{G}}$ and $\lambda_t^{\text{C}}$ represent the electricity retail price and the carbon market price at time $t$, respectively. $P_{t}^{\text{ch}}$ and $P_{t}^{\text{dis}}$ are the \ac{ESS} charging and discharging powers. $P_t^{\text{G}}$ is the net power exchanges with the grid, i.e.: 

\begin{figure}[t]
    \centering
	\includegraphics[width=0.75\linewidth]{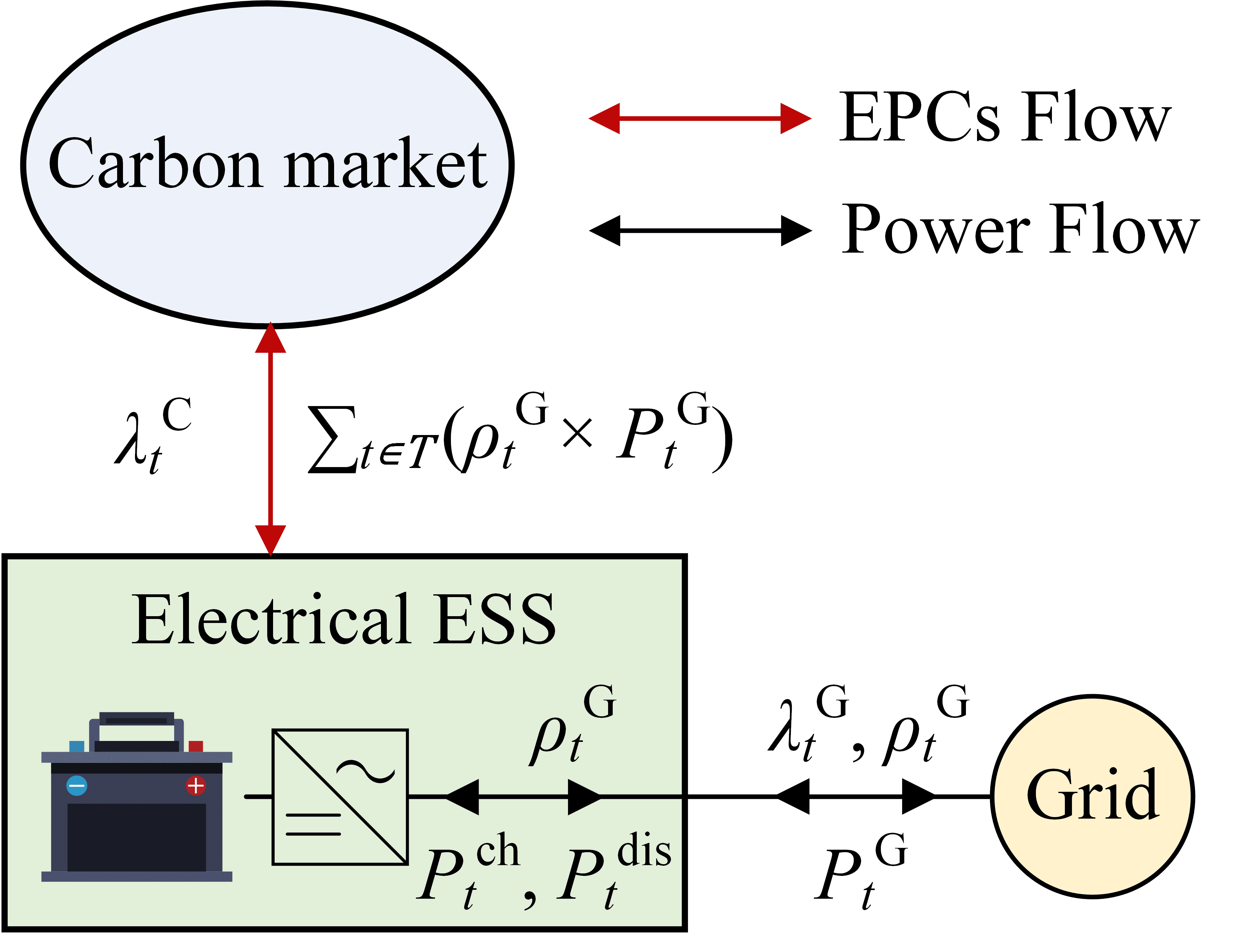}
	\caption{ \ac{ESS} carbon-aware dispatch structure.}
	\label{fig: Framework}
\end{figure}

\begin{equation}
\begin{split}
P_{t}^{\text{G}}=P_{t}^{\text{dis}} - P_{t}^{\text{ch}} \\
\forall t \in \mathcal{T}
\end{split}
\label{Balance}
\end{equation}

The optimal emission-aware operation of the \ac{ESS} is formulated as a single-stage optimization problem. A centralized \ac{EMS} determines the optimal \ac{ESS} dispatch. The objective function of \ac{ESS} operation includes both electricity cost savings and carbon revenue, i.e.: 

\begin{equation}
\begin{aligned}
\max_{P_t^{\text{G}}} \sum_{t \in \mathcal{T}}  
    \left( \lambda_{t}^{\text{G}} + \lambda_{t}^{\text{C}} \rho_{t}^{\text{G}} \right) P_{t}^{\text{G}}
\end{aligned}
\label{OPF_Cen}
\end{equation}

The \ac{ESS} operation is driven by real-time electricity prices, prevailing carbon market prices, and the grid’s \ac{MEI}. However, the \ac{ESS} introduces temporal coupling within the rolling-horizon optimization, thus influencing electricity procurement decisions across time steps. As a result, the dispatch schedule prioritizes charging \ac{ESS} when grid electricity is cheap and low in emissions; it favors discharging when grid electricity is carbon-intensive and costly. The \ac{ESS} charging and discharging powers are constrained by  (\ref{ES_Bound}). The \ac{ESS} state-of-charge (SoC) is modelled in (\ref{SOC}). 

\begin{equation}
\begin{split}
0<P_{t}^{\text{ch}}\le \bar{P}_{t}^{\text{ch}},0<P_{t}^{\text{dis}}\le\bar{P}_{t}^{\text{dis}} \\ \forall t \in \mathcal{T}
\end{split}
\label{ES_Bound}
\end{equation}

\begin{equation}
\begin{split}
0 \le \text{SoC}_{0}+\sum\limits_{s=0}^{t}{\left( \frac{\eta^{\text{ch}}P_{s}^{\text{ch}}}{E}-\frac{P_{s}^{\text{dis}}}{\eta^{\text{dis}}E} \right)} \le 1 \\
\; \forall t \in \mathcal{T}
\end{split}
\label{SOC}
\end{equation}

\noindent where $\bar{P}_{t}^{\text{ch}}$ and $\bar{P}_{t}^{\text{dis}}$ are the upper bound of \ac{ESS} charging/discharging power, with efficiencies $\eta^{\text{ch}}, \eta^{\text{dis}}$. $\text{SoC}_{0}$ is the initial SoC, and $E$ is the storage capacity.

\section{Results}
\subsection{Data}

\begin{figure}[t]
    \centering
	\includegraphics[width=\linewidth]{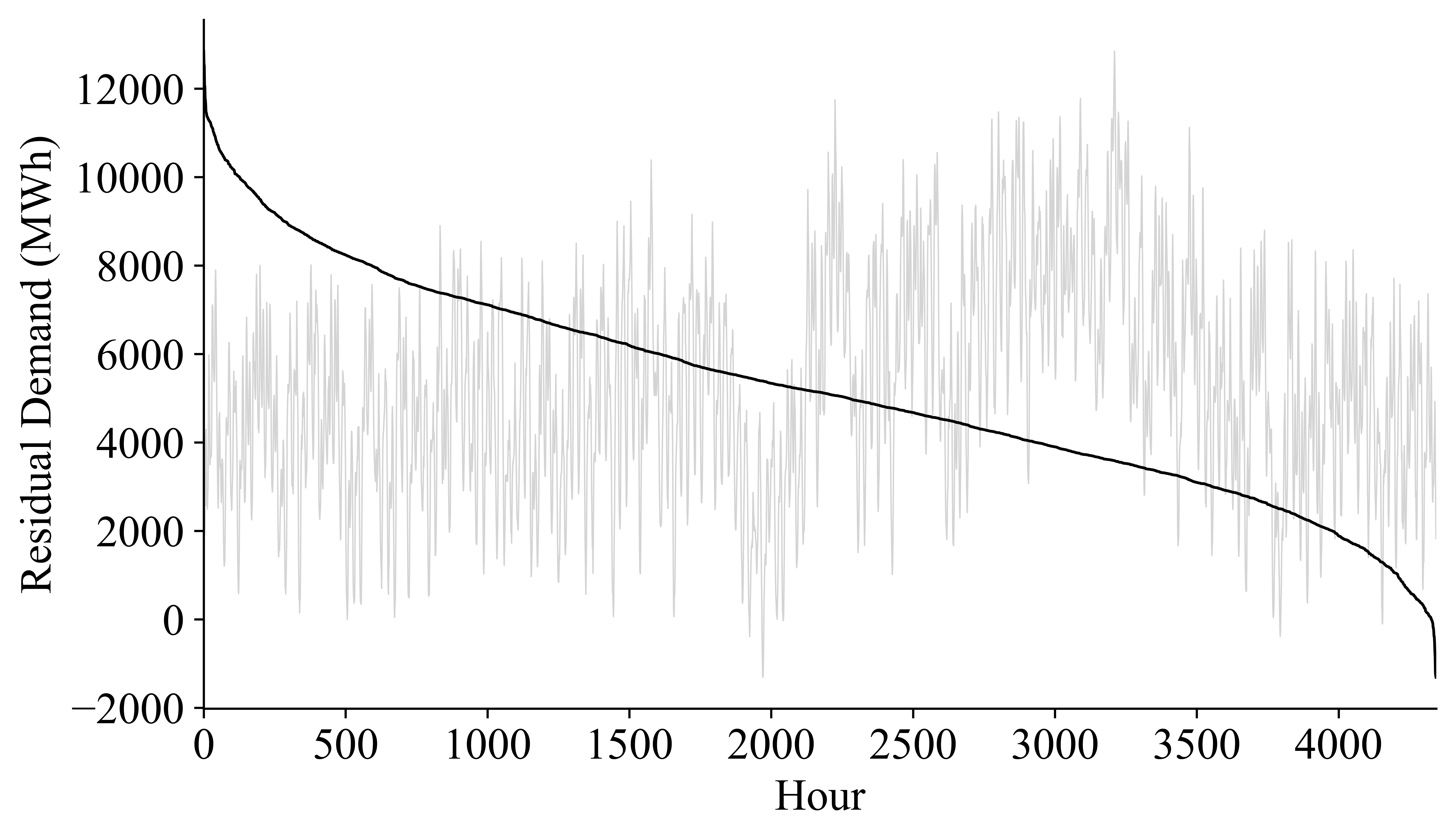}
	\caption{Hourly residual demand and its duration curve.}
	\label{fig: Res_Demand}
\end{figure}

\begin{figure}[t]
    \centering
	\includegraphics[width=\linewidth]{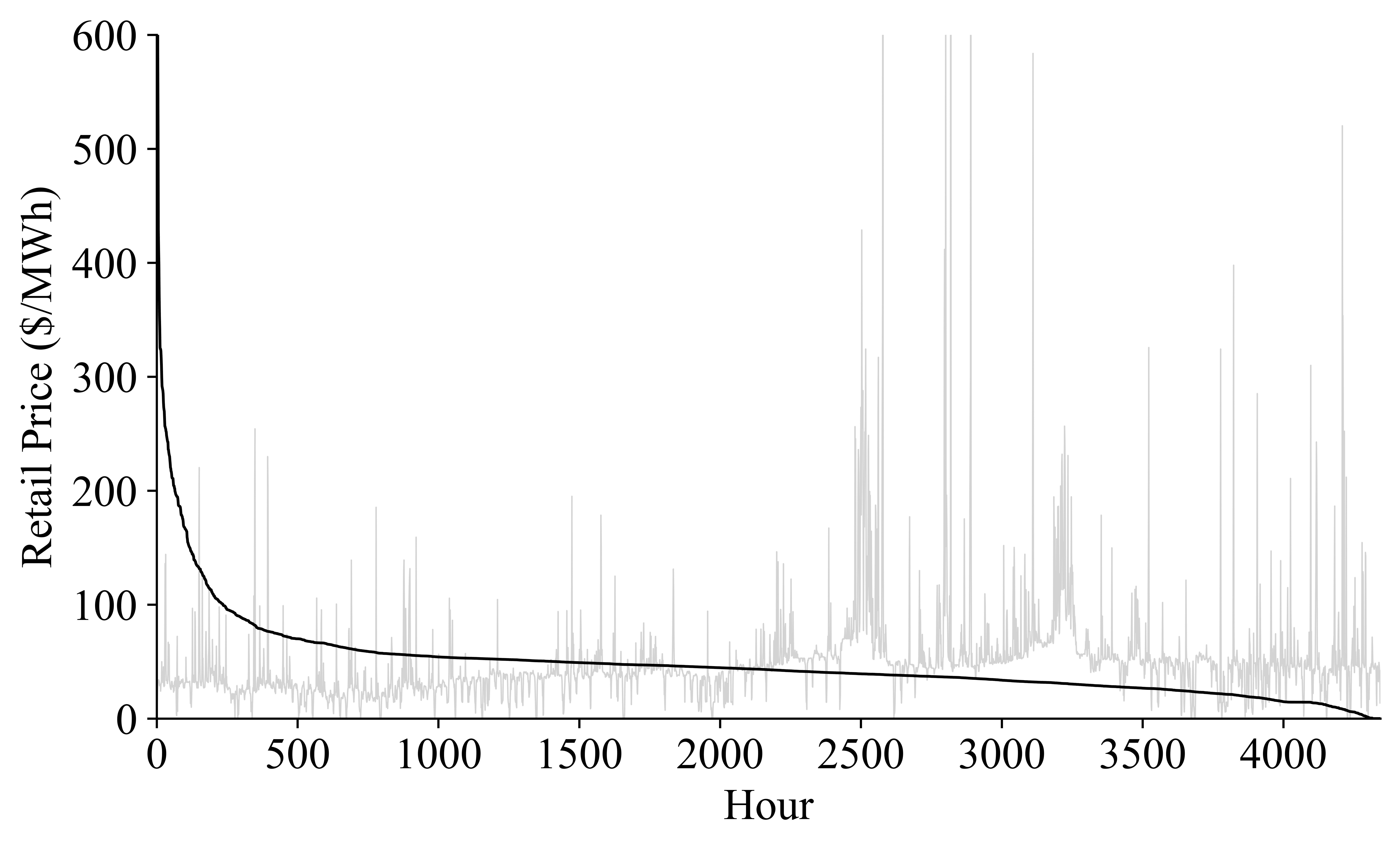}
	\caption{Hourly electricity price and its duration curve.}
	\label{fig: Retail_Price}
\end{figure}

For simulation purposes, this work considers data specific to the ecosystem of Ontario, Canada, including the residual demand and electricity retail price illustrated in Figures \ref{fig: Res_Demand} and \ref{fig: Retail_Price}, respectively (\cite{IESOReports}). These belong to an approximately 6-month period spanning 2024 and 2025. In these Figures, the grey lines represent the hourly data, and the black lines represent the duration curve sorted in descending order. Based on the federal carbon pollution pricing system of Canada, the carbon price is set as $\$80/\text{tCO}_2$ (\cite{ECCC}). The \ac{ESS} is modelled with a charging/discharging efficiency of 0.92. The dispatch of \ac{ESS} is based on the assumption that it has access to accurate predictions of electricity prices and \ac{MEI} over the optimization period. All simulations are modelled in Pyomo and solved by Gurobi on a Windows laptop with 13th Gen Intel(R) Core(TM) i9-13900H 2.60 GHz and 32 GB of RAM.

\subsection{Dispatch Comparisons}
To evaluate the proposed framework's performance, three case studies are designed, as follows:

\begin{itemize}
    \item \textbf{Case 1}: In this case, the \ac{ESS} operates solely considering the electricity price.
    \item \textbf{Case 2}: In this case, the \ac{ESS} operates solely considering the carbon price.
    \item \textbf{Case 3}: In this case, the \ac{ESS} responds to both carbon and electricity prices, reflecting a dual-cost and emission-aware strategy.
\end{itemize}

\begin{figure*}[t]
    \centering

    \begin{subfigure}[b]{0.48\textwidth}
        \includegraphics[width=\linewidth]{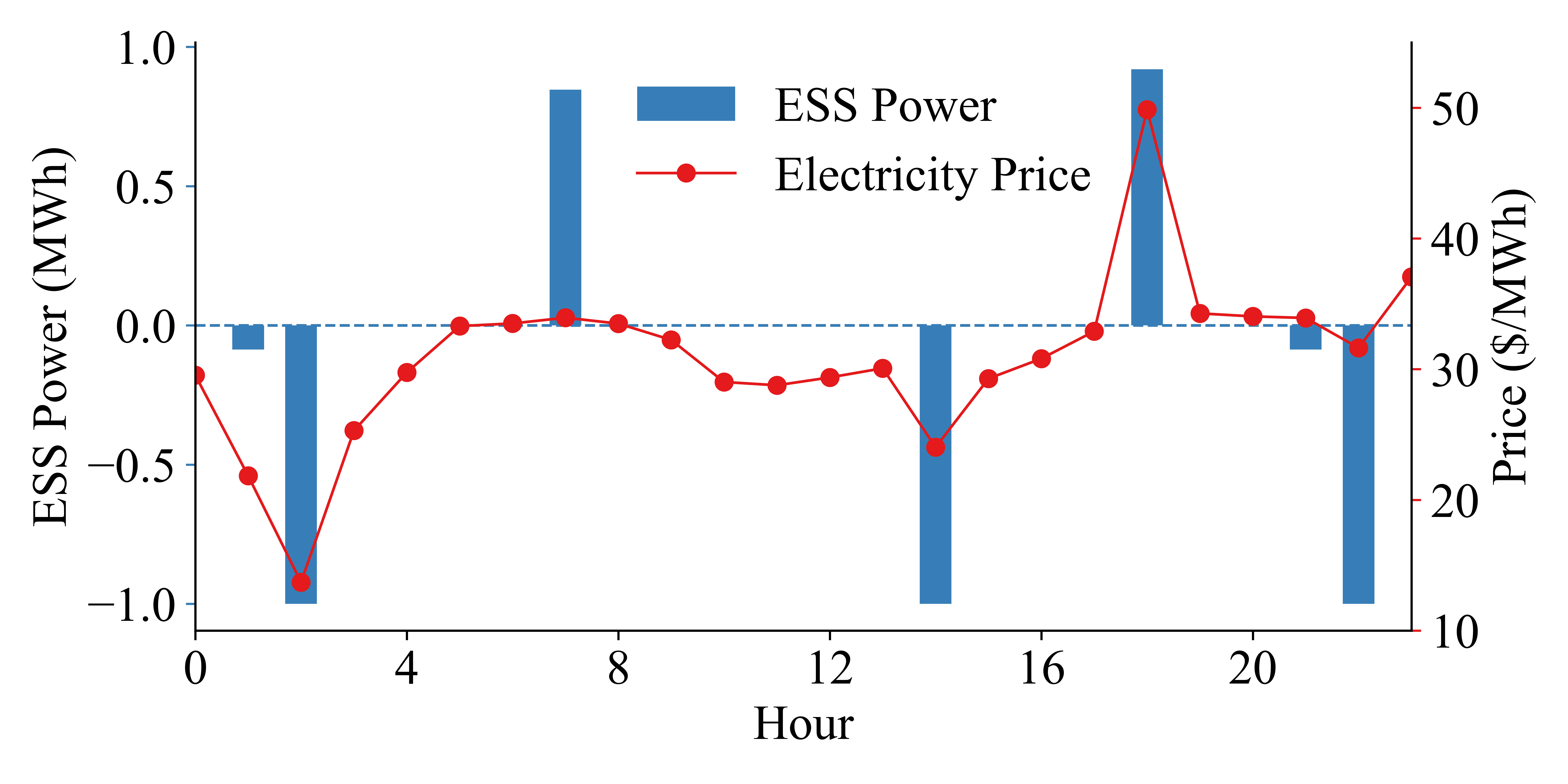}
        \caption{Case 1 with 1MWh of ESS}
        \label{fig:BP1}
    \end{subfigure}
    \hfill
    \begin{subfigure}[b]{0.48\textwidth}
        \includegraphics[width=\linewidth]{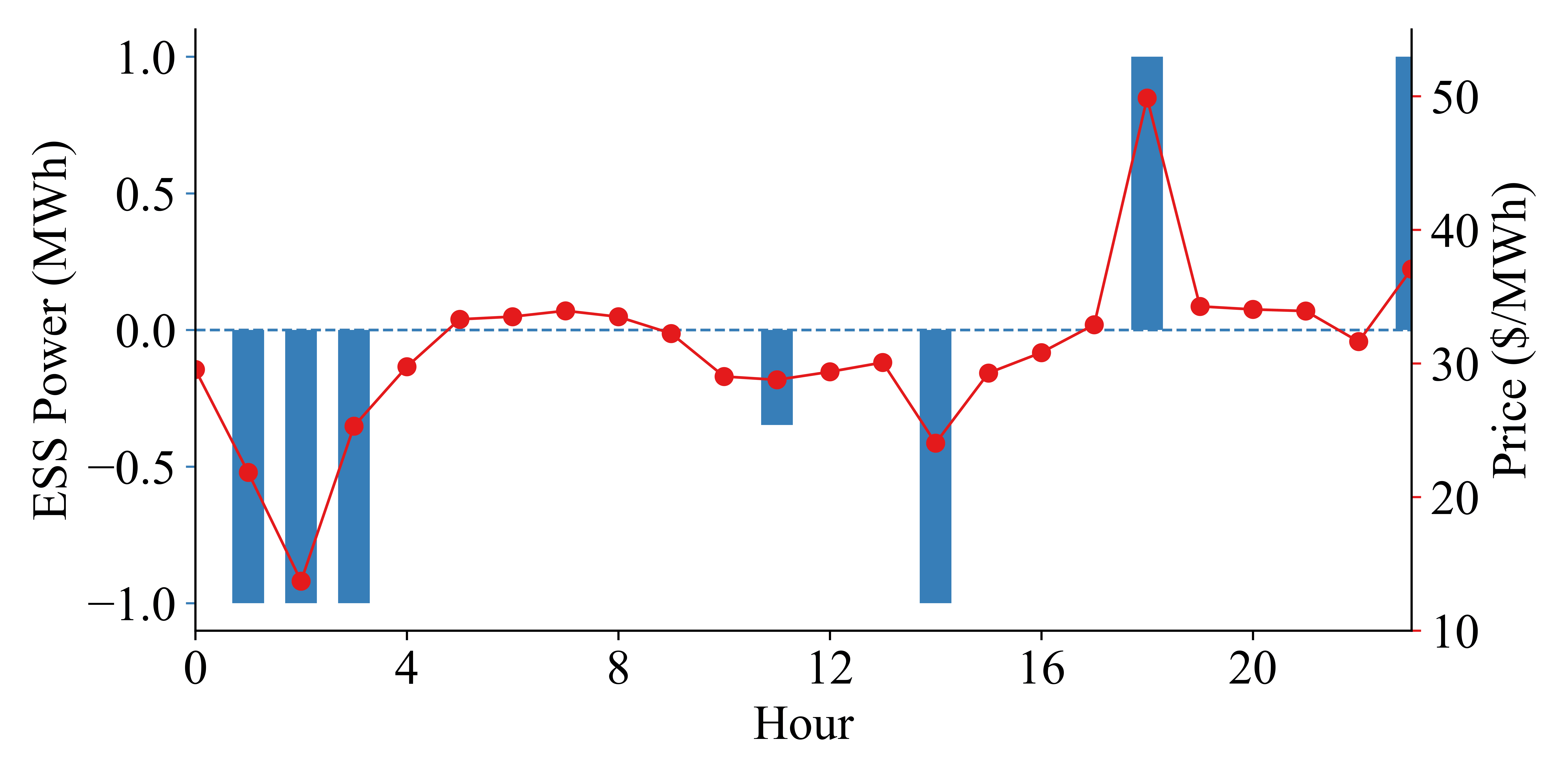}
        \caption{Case 1 with 4MWh of ESS}
        \label{fig:BP2}
    \end{subfigure}

    \begin{subfigure}[b]{0.48\textwidth}
        \includegraphics[width=\linewidth]{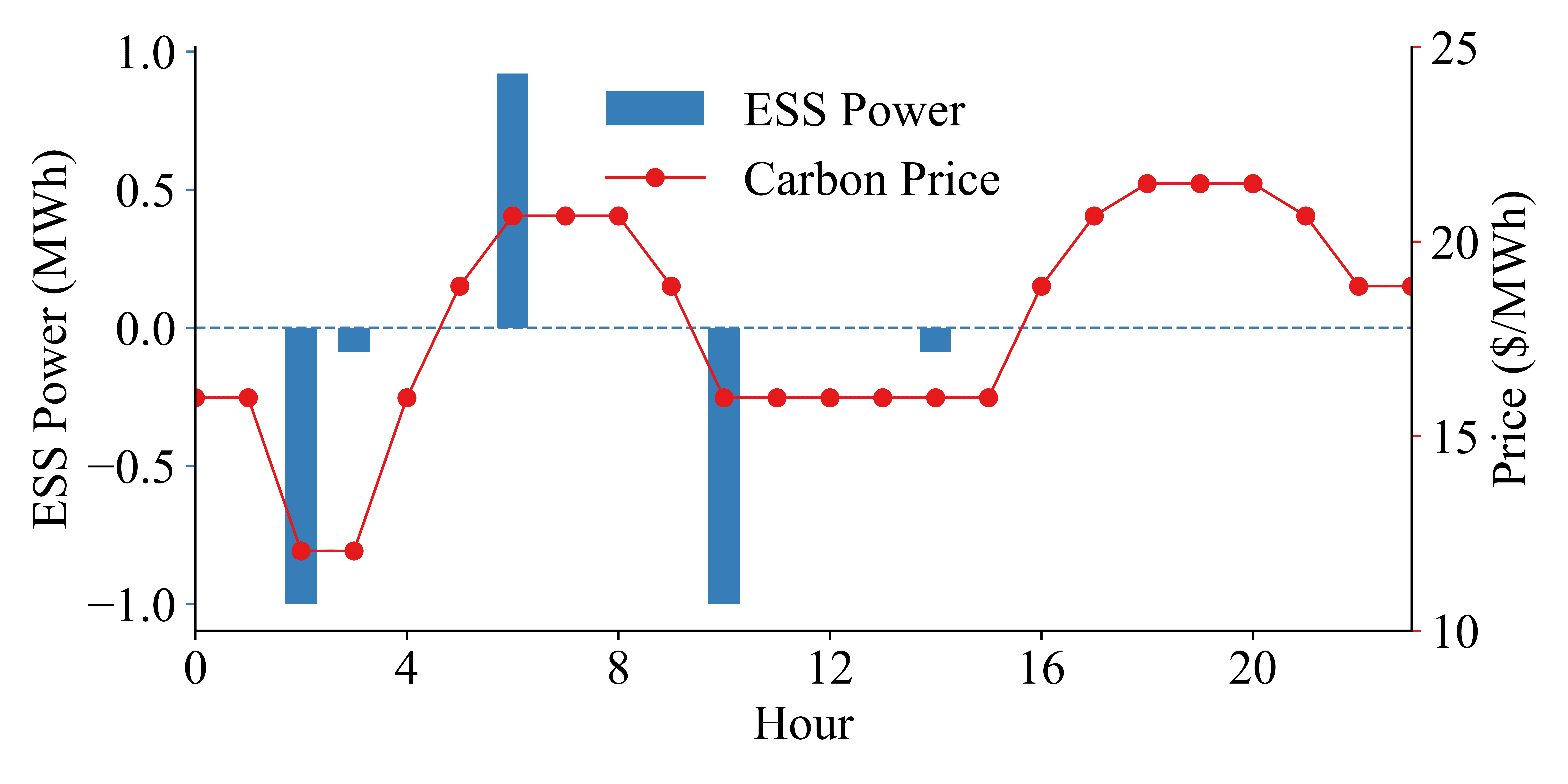}
        \caption{Case 2 with 1MWh of ESS}
        \label{fig:BP3}
    \end{subfigure}
    \hfill
    \begin{subfigure}[b]{0.48\textwidth}
        \includegraphics[width=\linewidth]{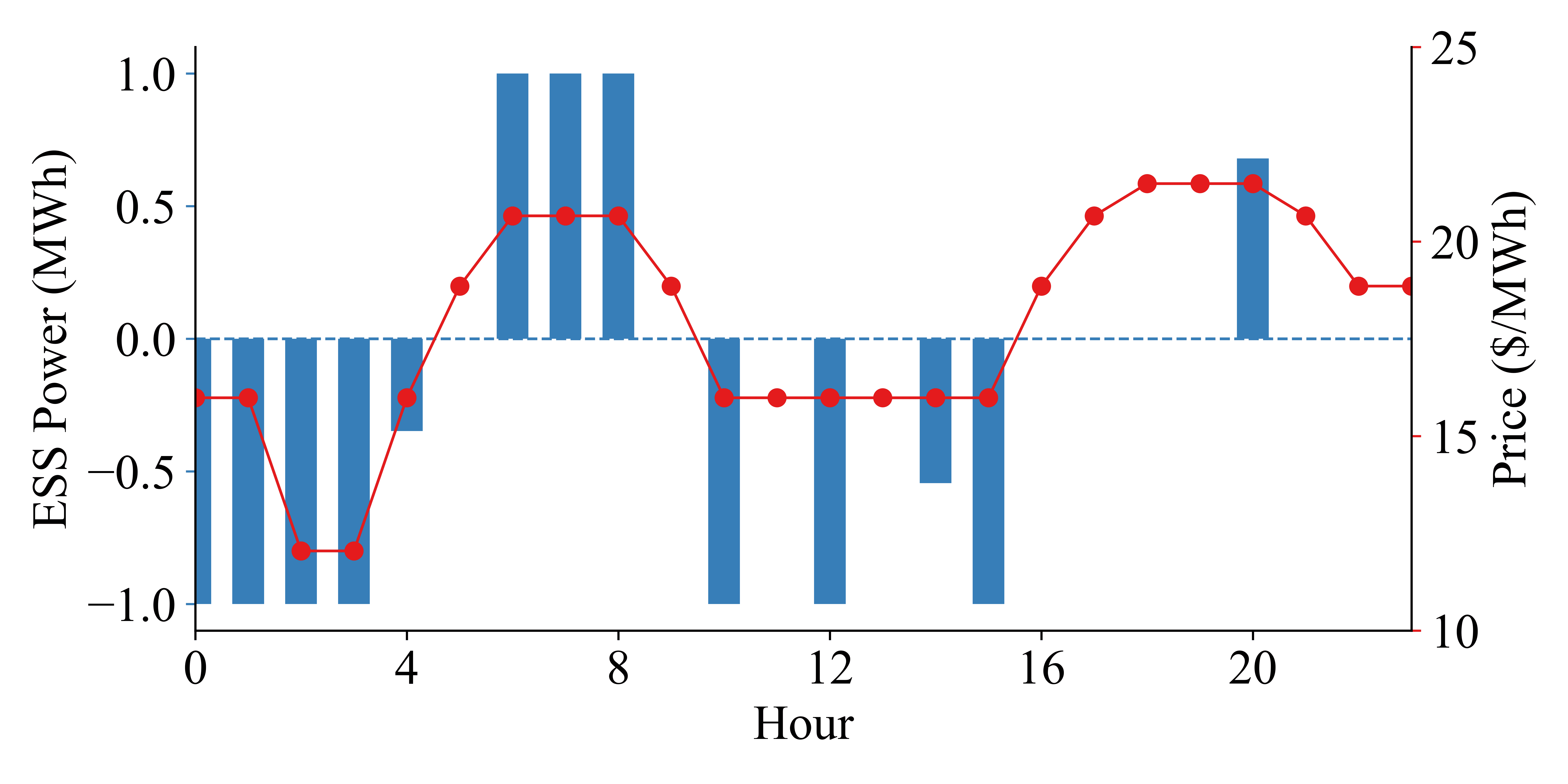}
        \caption{Case 2 with 4MWh of ESS}
        \label{fig:BP4}
    \end{subfigure}

    \begin{subfigure}[b]{0.48\textwidth}
        \includegraphics[width=\linewidth]{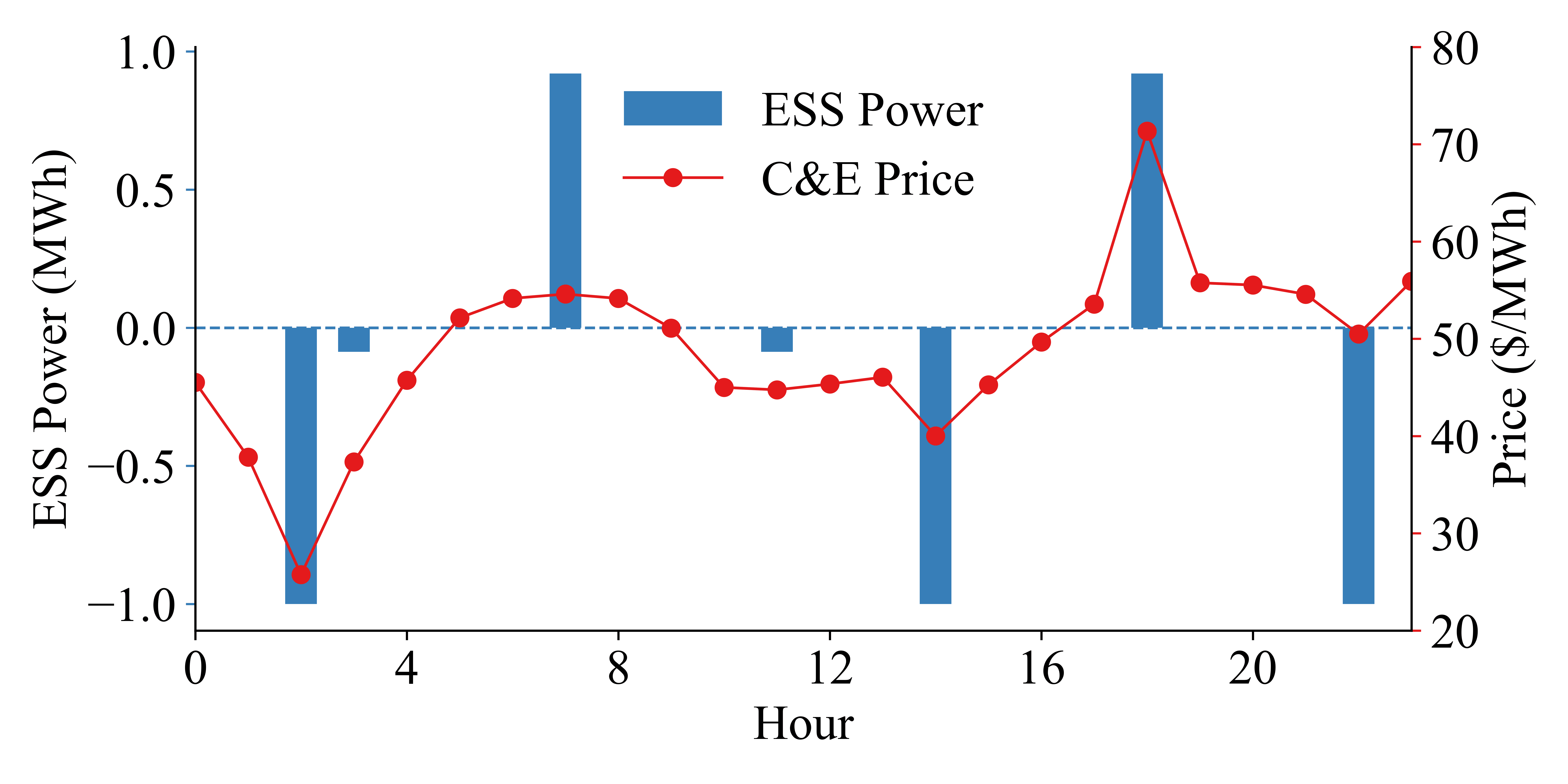}
        \caption{Case 3 with 1MWh of ESS}
        \label{fig:BP5}
    \end{subfigure}
    \hfill
    \begin{subfigure}[b]{0.48\textwidth}
        \includegraphics[width=\linewidth]{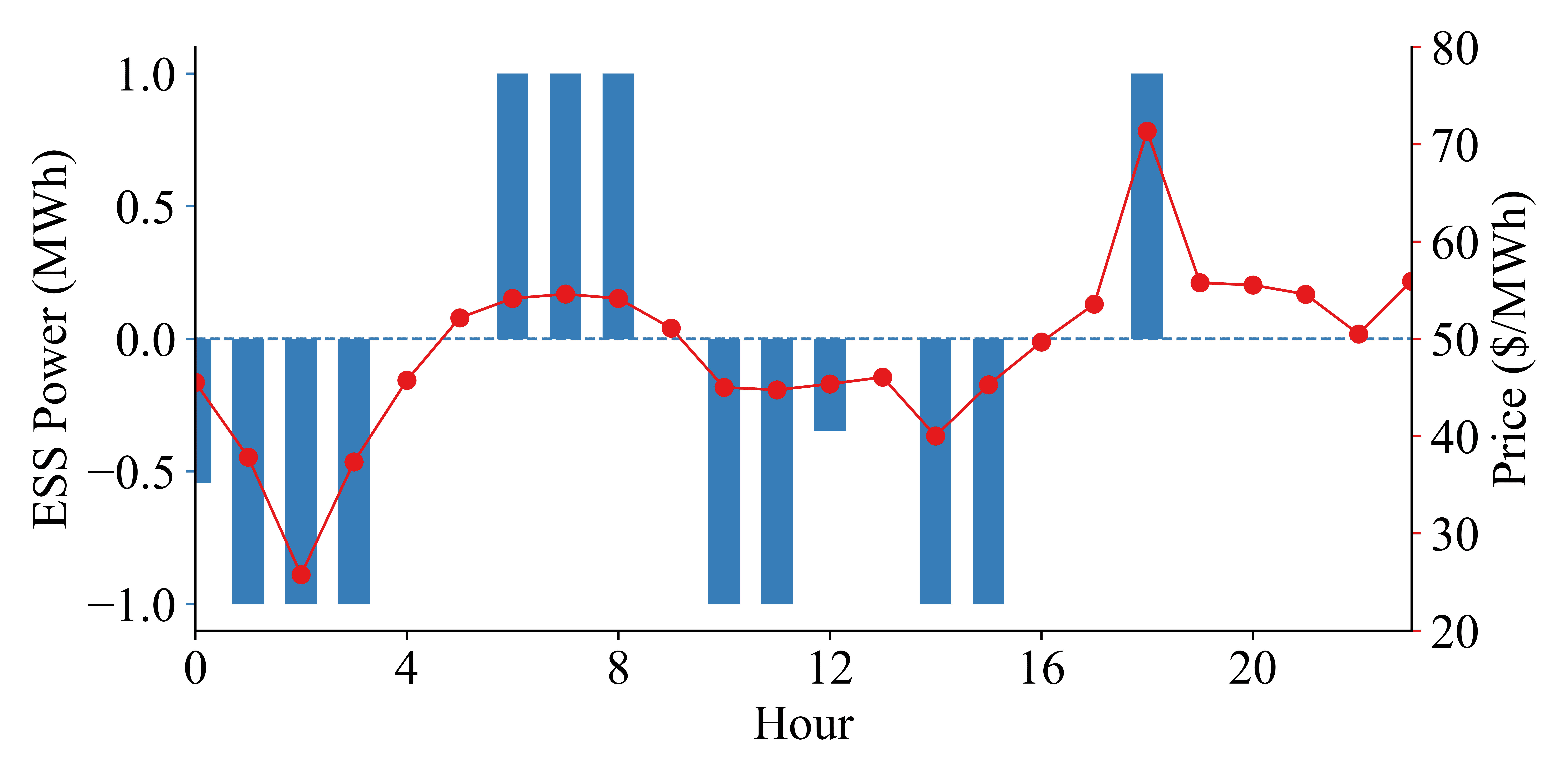}
        \caption{Case 3 with 4MWh of ESS}
        \label{fig:BP6}
    \end{subfigure}

    \caption{\ac{ESS} dispatch across three pricing cases with different \ac{ESS} capacity.}
    \label{fig:ESS_Price}
\end{figure*}

The \ac{ESS} dispatch results for the above cases are reported in Figures \ref{fig:ESS_Price}. The carbon price shown in this figure is calculated as $\lambda_{t}^{\text{C}} \rho _{t}^{\text{G}}$ in \$/MWh. Moreover, the price reported for Case 3 is the summation of both the carbon and electricity prices. For each scenario, the analysis is carried out for two \ac{ESS} capacities, 1 MWh and 4 MWh, both with a maximum charging and discharging power of 1 MW. It is observed that within each case, the 1 MWh \ac{ESS} exhibits a lower dispatch frequency compared to the 4 MWh system, despite sharing the same power limit. This is due to its lower energy capacity, which constrains the duration and depth of charge/discharge cycles. Consequently, the 1 MWh battery must prioritize only the most economically or environmentally favorable periods. In contrast, the 4 MWh \ac{ESS} offers greater flexibility, allowing for more frequent responses to pricing signals, thereby unlocking higher value and aligning more effectively with system objectives. Note that the overall performance strategy is similar across the three cases for the 1 MWh and 4 MWh \ac{ESS}. The following analysis will therefore focus on the 4 MWh  \ac{ESS} across the three cases.

For the conventional operation conditions, i.e.. Case 1, it can be observed that the \ac{ESS} operation may lead to increased operational carbon footprint. A comparison with Case 2, in which the ESS operates greedily on emission reduction, highlight this drawback more effectively. For instance, as shown in Figures~\ref{fig:BP4} and~\ref{fig:BP6}, during 6 AM–8 AM, Case 2 discharges in response to high carbon intensity, while Case 1 remains inactive due to relatively stable electricity prices, highlighting the environmental blind spots of traditional market-based dispatch strategies. On the other hand, while Case 2 leads to an effective carbon reduction behavior, it may result in missed economic opportunities by discharging or remaining inactive during low-price periods. This trade-off is evident between 6 PM-8 PM, where the \ac{ESS} operation diverges across the cases. In Case 2, the \ac{ESS} discharges around 8 PM to target a period of high \ac{MEI}. In contrast, Cases 1 and 3 shift it to 6 PM, aligning instead with elevated electricity prices. In this context, Case 3 demonstrates a balanced strategy that considers both the carbon and electricity prices, as shown in Figures \ref{fig:BP2}. Thus, it can be observed that the \ac{ESS} discharges during periods when the combined prices are high and charges during periods with a low electricity price and less intensive operational emissions. 

\begin{figure}[t]
    \centering
	\includegraphics[width=\linewidth]{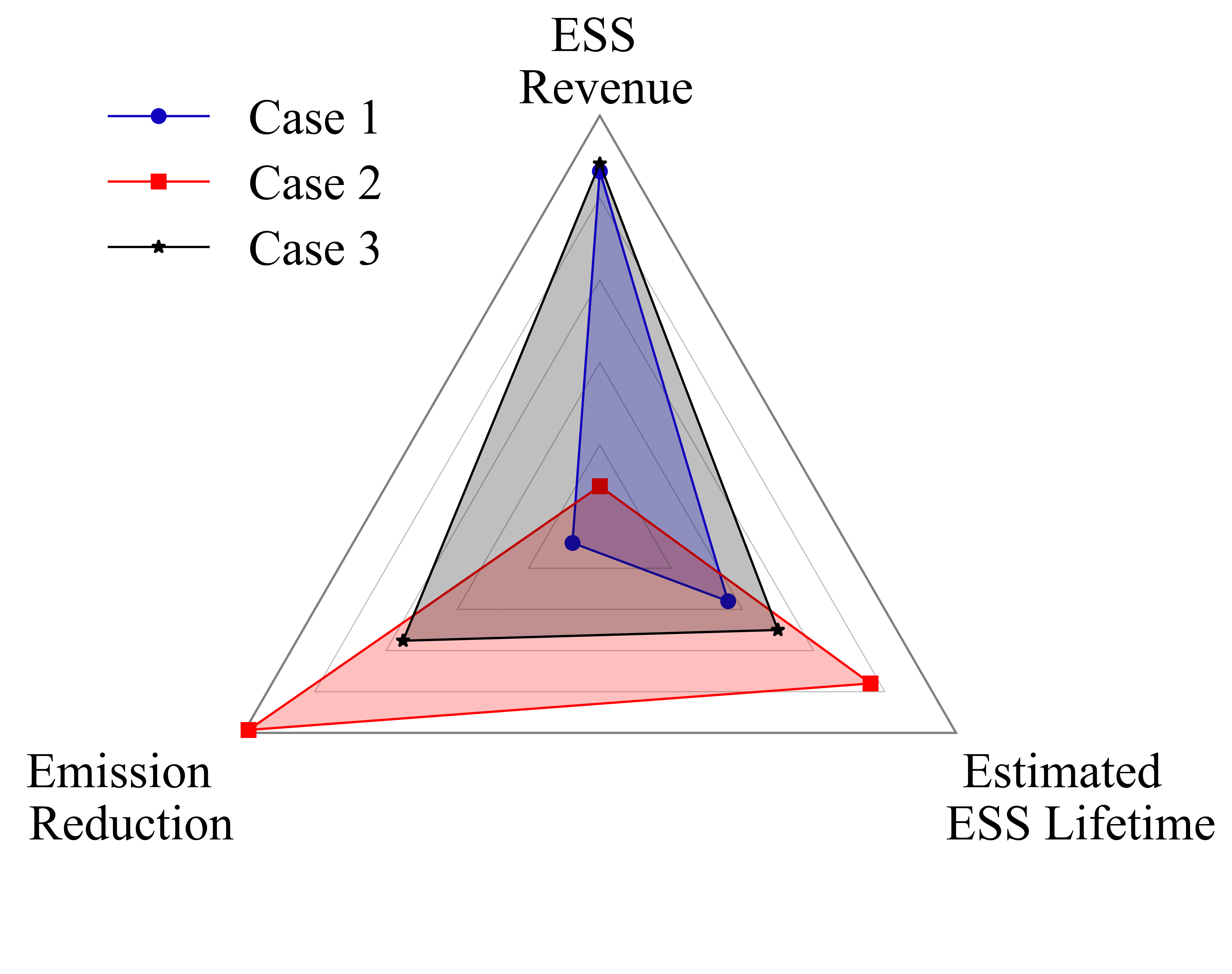}
	\caption{Normalized average performance (wider better).}
	\label{fig: radar}       
\end{figure}

To quantitatively compare the performance of the \ac{ESS} within and across the three cases, Table \ref{Num_Results} report the numerical values for operational revenue and emission reduction. Moreover, while the operational impact on the \ac{ESS} lifetime is not considered in the optimization formulations, Table \ref{Num_Results} also report the \ac{ESS} remaining lifetime to increase the operational impact visibility \footnote{The \ac{ESS} remaining lifetime is quantified as one minus the ratio of accumulated full equivalent charge-discharge cycles to the \ac{ESS}’s rated cycle life, assumed to be 3,000 cycles in this study.}. Thus, it can be observed that Case 3 demonstrates the most balanced performance, achieving the highest revenue while maintaining moderate emission reduction and battery longevity, highlighting the value of a combined carbon-aware and market-responsive dispatch strategy. Case 2 achieves the most substantial emission reductions and the longest \ac{ESS} operational lifetime, as it is driven solely by carbon intensity signals; however, this comes at the expense of economic profitability. In contrast, Case 1 achieves moderate revenue but offers minimal emission benefits and the shortest lifetime, as it responds only to electricity prices. These results illustrate that integrating both carbon and electricity pricing signals enables \ac{ESS} operation that is economically viable, environmentally responsible, and technically sustainable. Finally, to provide a general qualitative comparison, the normalized average performances among the three cases for the 1 MWh \ac{ESS} are shown in Figure \ref{fig: radar}. While it is not accurate to directly compare the length of dimensions against one another, the surface area of each case provides a relative indication of the operational welfare.

\begin{table}[t]
\centering
\caption{Numerical results for three cases.}
\begin{tabular}{l|>{\centering\arraybackslash}p{1.5cm}>{\centering\arraybackslash}p{1.5cm}>{\centering\arraybackslash}p{1.5cm}}
\hline
Case & Operational Revenue ($10^3\$$) & Emission Reduction ($\text{tCO}_2$) & \ac{ESS} remaining lifetime\\
\hline
1  & 17.30 & -6.62 & 0.68 \\
2  &  1.99 & 9.75 & 0.88 \\
3  &  17.67 & 1.94 & 0.75 \\
\hline
\end{tabular}
\label{Num_Results}
\end{table}

\subsection{Sensitivity Analysis}

\begin{figure}[t]
    \centering

    \begin{subfigure}[b]{0.45\textwidth}
        \includegraphics[width=\linewidth]{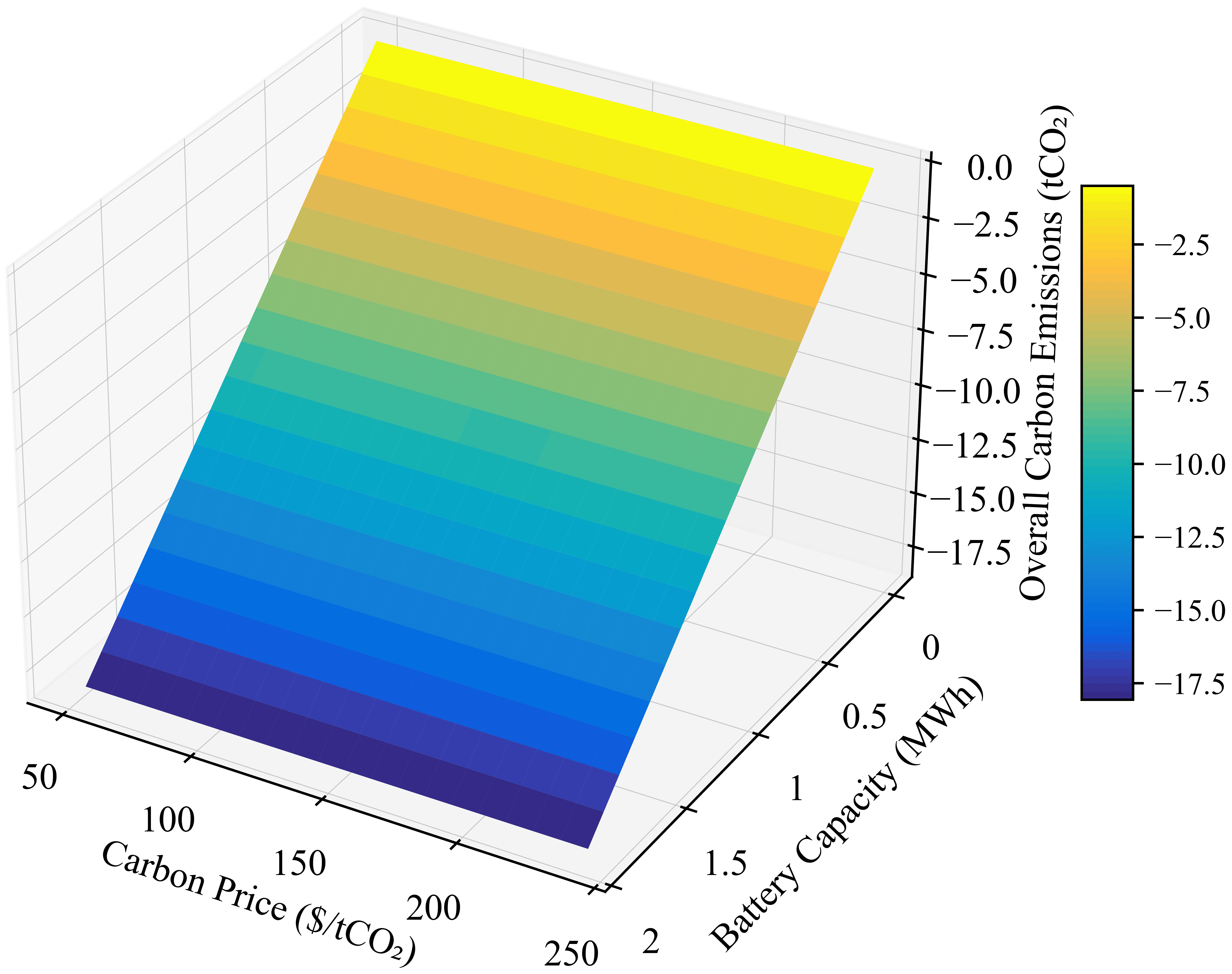}
        \caption{Case 2: Carbon price only}
        \label{fig:3D2}
    \end{subfigure}
    
    \begin{subfigure}[b]{0.45\textwidth}
        \includegraphics[width=\linewidth]{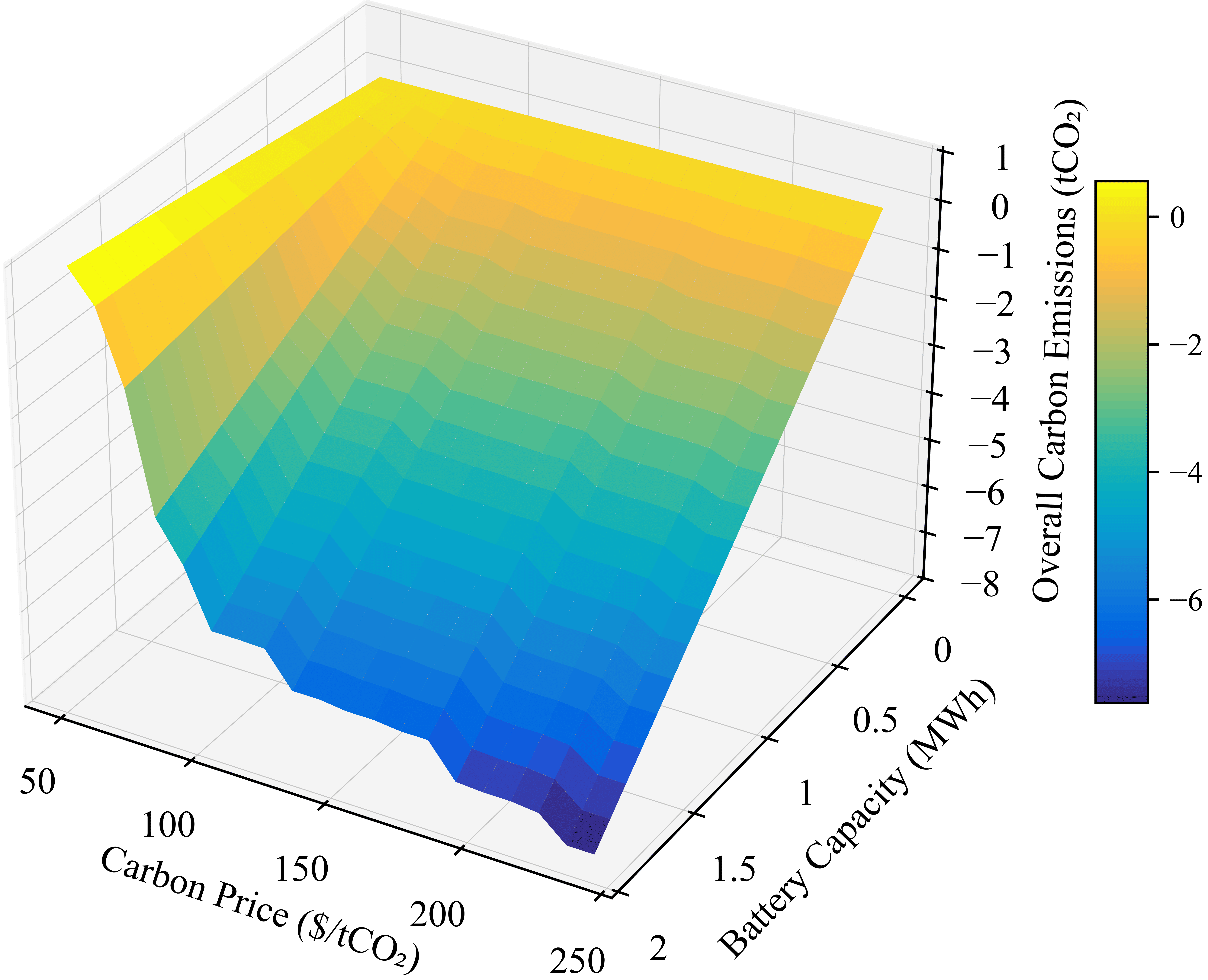}
        \caption{Case 3: Carbon and electricity prices}
        \label{fig:3D1}
    \end{subfigure}


    \caption{Impact of \ac{ESS} capacity and carbon price on overall carbon emissions across Cases 2 and 3.}
    \label{fig:3D}
\end{figure}

Figure~\ref{fig:3D} illustrates the overall carbon emissions as a function of \ac{ESS} capacity and carbon price for Cases 2 and 3, assuming the \ac{ESS} with a C-rate of 1. Note that no results are reported for case 1 since it is insensitive to the carbon price. As shown in Figure \ref{fig:3D2}, in Case 2, emissions consistently decrease as \ac{ESS} capacity increases. This indicates that greater storage flexibility improves the system’s ability to shift consumption away from high-emission periods. The emission surface remains flat along the carbon price axis, suggesting that in the absence of electricity price influence, the mere presence of a carbon signal, rather than its magnitude, is sufficient to trigger carbon-aware dispatch. On the other hand, in Case 3, as shown in Figure \ref{fig:3D1}, emissions decrease most significantly as both \ac{ESS} capacity and carbon price increase. While the decrease in emissions remains relatively modest across each axis alone, it decreases aggressively along the diagonal axis, where both variables increase together. This suggests that the most effective operational strategy is obtained where the \ac{ESS} capacity is proportionally designed for a jurisdiction's carbon pricing; a higher carbon price incentivizes a higher investment in \ac{ESS} capacity. Together, these figures highlight the importance of coordinating fully cost-reflective pricing strategies and \ac{ESS} capacity to achieve an economical decarbonization.

\section{Conclusion and Future Works} 

This paper presents an emission-aware operation framework for \ac{ESS}, agnostic to their size or operational voltage. It is built on a proposed real-time grid MEI calculation based on real-world data and considering the grid's operation strategy. Thus, the framework enables the \ac{ESS} across transmission and distribution systems to participate in emission reduction mandates by generating tradable \ac{EPCs}, effectively transforming \ac{ESS} into carbon-responsive assets. It is shown through numerical studies, based on real-world data, that the proposed strategy delivers an emission-aware operation while enhancing the operational revenue. A major difference of the proposed framework compared to other dual emission-cost optimization approaches is the explicit quantification of the emission costs tied to the practicalities of the grid control, eliminating the need for compromised optimization with arbitrary tunable coefficients. Simulation results further reveal that substantial emissions reduction is achieved only when both carbon pricing and \ac{ESS} capacity are scaled concurrently. Increasing either factor in isolation yields limited benefits, highlighting the importance of coordinated policy and infrastructure planning. Overall, the proposed framework offers a scalable, transparent, and policy-aligned pathway for \ac{ESS} of all sizes and deployment levels to unlock carbon revenue streams through participation in compliance carbon markets.

Future research will investigate the influence of network constraints on \ac{ESS} dispatch and explore distributed coordination mechanisms for independent \ac{ESS} units. Additionally, the emission accounting methodology should be refined by disaggregating \ac{MEI} into locational marginal emissions, allowing for more spatially precise and equitable coordination of carbon-aware \ac{ESS} operations across the power system.

\textbf{Git Repo}: Data and codes will be publicly available upon acceptance.



\printbibliography

\end{document}

%% file: hicss-packages.tex
\usepackage[letterpaper]{geometry}
\usepackage{hicss}
\usepackage{times}
\usepackage[none]{hyphenat}
\usepackage{url}
\usepackage{latexsym}
\usepackage{indentfirst}
\usepackage{graphicx}
\graphicspath{{images/}}
\usepackage[
    style=apa,
  ]{biblatex}

%% file: hicss.bib
@misc{government2019technology,
  title={{Technology innovation and emissions reduction regulation}},
  author={{Government of Alberta}},
  year={2023},
url = {https://www.alberta.ca/technology-innovation-and-emissions-reduction-regulation},
  publisher={Climate Change and Emissions Management Act Edmonton, AB}
}

@misc{ElectricityMaps,
  author    = "{Electricity Maps}",
  title     = "{{Electricity Grid Carbon Emissions}}",
  year      = "2025",
  url       = "https://app.electricitymaps.com/",
  note      = "Online"
}

@article{he2024locational,
  title={{Is locational marginal price all you need for locational marginal emission?}},
  author={He, Xuan and Tsang, Danny HK and Chen, Yize},
  journal={arXiv preprint arXiv:2411.12104},
  year={2024}
}

@ARTICLE{Jiang20234724,
	author = {Jiang, Kai and Liu, Nian and Yan, Xiaohe and Xue, Yusheng and Huang, Jie},
	title = {{Modeling strategic behaviors for GenCo with joint consideration on electricity and carbon markets}},
	year = {2023},
	journal = {IEEE Transactions on Power Systems},
	volume = {38},
	number = {5},
	pages = {4724 – 4738},
	type = {Article},
	publication_stage = {Final},
	source = {Scopus},
}

@ARTICLE{Gu2023790,
	author = {Gu, Chenjia and Liu, Yikui and Wang, Jianxue and Li, Qingtao and Wu, Lei},
	title = {{Carbon-oriented planning of distributed generation and energy storage assets in power distribution network with hydrogen-based microgrids}},
	year = {2023},
	journal = {IEEE Transactions on Sustainable Energy},
	volume = {14},
	number = {2},
	pages = {790 – 802},
	type = {Article},
	publication_stage = {Final},
	source = {Scopus}
}

@misc{IESOReports,
 title={IESO Public Reports},
 author={{IESO}},
year={2025},
 url={https://reports-public.ieso.ca/public/},
 note={(Accessed: 30 April, 2025)},
 }

@misc{ECCC,
  title = {The federal carbon pollution pricing benchmark},
 author = {{Environment and Climate Change Canada}},
 url={https://www.canada.ca/en/environment-climate-change.html},
  year={2025},
}

@ARTICLE{9770947,
  author={Mu, Chenggang and Ding, Tao and Zhu, Shanying and Han, Ouzhu and Du, Pengwei and Li, Fangxing and Siano, Pierluigi},
  journal={IEEE Transactions on Smart Grid}, 
  title={A Decentralized Market Model for a Microgrid With Carbon Emission Rights}, 
  year={2023},
  volume={14},
  number={2},
  pages={1388-1402},
 }

@article{piperagkas2011stochastic,
  title={Stochastic PSO-based heat and power dispatch under environmental constraints incorporating CHP and wind power units},
  author={Piperagkas, GS and Anastasiadis, AG and Hatziargyriou, ND},
  journal={Electric Power Systems Research},
  volume={81},
  number={1},
  pages={209--218},
  year={2011},
  publisher={Elsevier}
}

@misc{EUETS,
 title={EU Emissions Trading System},
 author = {{European Union}},
 url={https://climate.ec.europa.eu/eu-action/eu-emissions-trading-system-eu-ets_en},
  year={2023},
 }

@ARTICLE{7999305,
  author={Melhem, Fady Y. and Grunder, Olivier and Hammoudan, Zakaria and Moubayed, Nazih},
  journal={Canadian Journal of Electrical and Computer Engineering}, 
  title={Optimization and Energy Management in Smart Home Considering Photovoltaic, Wind, and Battery Storage System With Integration of Electric Vehicles}, 
  year={2017},
  volume={40},
  number={2},
  pages={128-138},
 }

@inproceedings{azuatalam2018techno,
  title={Techno-economic Analysis of Residential PV-battery Self-consumption},
  author={Azuatalam, Donald and F{\"o}rstl, Markus and Paridaric, Kaveh and Ma, Yiju and Chapman, Archie C and Verbi{\v{c}}, Gregor},
  booktitle={Asia-Pacific Solar Research Conference 2018},
  year={2018},
  organization={Australian PV Institute}
}

@misc{IEA_BESS,
 title={Batteries and Secure Energy Transitions},
 author = {{International Energy Agency}},
 url={https://www.iea.org/reports/batteries-and-secure-energy-transitions},
  year={2024},
 }

@ARTICLE{9740444,
  author={Abomazid, Abdulrahman M. and El-Taweel, Nader A. and Farag, Hany E. Z.},
  journal={IEEE Transactions on Sustainable Energy}, 
  title={Optimal Energy Management of Hydrogen Energy Facility Using Integrated Battery Energy Storage and Solar Photovoltaic Systems}, 
  year={2022},
  volume={13},
  number={3},
  pages={1457-1468},
 }

@ARTICLE{9166729,
  author={Bhattacharjee, Shubhrajit and Sioshansi, Ramteen and Zareipour, Hamidreza},
  journal={IEEE Transactions on Power Systems}, 
  title={Benefits of Strategically Sizing Wind-Integrated Energy Storage and Transmission}, 
  year={2021},
  volume={36},
  number={2},
  pages={1141-1151},
 }

@ARTICLE{8003298,
  author={Carrión, Miguel and Dvorkin, Yury and Pandžić, Hrvoje},
  journal={IEEE Transactions on Power Systems}, 
  title={Primary Frequency Response in Capacity Expansion With Energy Storage}, 
  year={2018},
  volume={33},
  number={2},
  pages={1824-1835},
 }

@ARTICLE{9557813,
  author={Steriotis, Konstantinos and Šepetanc, Karlo and Smpoukis, Konstantinos and Efthymiopoulos, Nikolaos and Makris, Prodrommos and Varvarigos, Emmanouel and Pandžić, Hrvoje},
  journal={IEEE Transactions on Sustainable Energy}, 
  title={Stacked Revenues Maximization of Distributed Battery Storage Units Via Emerging Flexibility Markets}, 
  year={2022},
  volume={13},
  number={1},
  pages={464-478},
 }

@ARTICLE{9089020,
  author={Haghighat, Hossein and Karimianfard, Hossein and Zeng, Bo},
  journal={IEEE Transactions on Smart Grid}, 
  title={Integrating Energy Management of Autonomous Smart Grids in Electricity Market Operation}, 
  year={2020},
  volume={11},
  number={5},
  pages={4044-4055},
 }

@article{zafirakis2015embodied,
  title={Embodied CO2 emissions and cross-border electricity trade in Europe: Rebalancing burden sharing with energy storage},
  author={Zafirakis, Dimitrios and Chalvatzis, Konstantinos J and Baiocchi, Giovanni},
  journal={Applied Energy},
  volume={143},
  pages={283--300},
  year={2015},
  publisher={Elsevier}
}

@ARTICLE{8844848,
  author={Olsen, Daniel J. and Kirschen, Daniel S.},
  journal={IEEE Transactions on Power Systems}, 
  title={Profitable Emissions-Reducing Energy Storage}, 
  year={2020},
  volume={35},
  number={2},
  pages={1509-1519},
 }

@article{hittinger2015bulk,
  title={Bulk energy storage increases United States electricity system emissions},
  author={Hittinger, Eric S and Azevedo, In{\^e}s ML},
  journal={Environmental science \& technology},
  volume={49},
  number={5},
  pages={3203--3210},
  year={2015},
  publisher={ACS Publications}
}

@article{babacan2018unintended,
  title={Unintended effects of residential energy storage on emissions from the electric power system},
  author={Babacan, Oytun and Abdulla, Ahmed and Hanna, Ryan and Kleissl, Jan and Victor, David G},
  journal={Environmental science \& technology},
  volume={52},
  number={22},
  pages={13600--13608},
  year={2018},
  publisher={ACS Publications}
}

@article{lin2016emissions,
  title={Emissions impacts of using energy storage for power system reserves},
  author={Lin, Yashen and Johnson, Jeremiah X and Mathieu, Johanna L},
  journal={Applied energy},
  volume={168},
  pages={444--456},
  year={2016},
  publisher={Elsevier}
}

@article{feng2022bi,
  title={Bi-level optimal capacity planning of load-side electric energy storage using an emission-considered carbon incentive mechanism},
  author={Feng, Jieran and Zhou, Hao},
  journal={Energies},
  volume={15},
  number={13},
  pages={4592},
  year={2022},
  publisher={MDPI}
}

@misc{colbert2021greenhouse,
  title={Greenhouse Gas Emissions Accounting for Battery Energy Storage Systems (BESS)},
  author={Colbert-Sangree, T and Gillenwater, M and Diamant, A and Fischer, L},
  year={2021},
  publisher={EPRI Palo Alto, CA}
}

@article{du2024real,
  title={Real-time energy management for net-zero power systems based on shared energy storage},
  author={Du, Pengbo and Huang, Bonan and Liu, Ziming and Yang, Chao and Sun, Qiuye},
  journal={Journal of Modern Power Systems and Clean Energy},
  volume={12},
  number={2},
  pages={371--380},
  year={2024},
  publisher={SGEPRI}
}

@ARTICLE{7021901,
  author={Kang, Chongqing and Zhou, Tianrui and Chen, Qixin and Wang, Jianhui and Sun, Yanlong and Xia, Qing and Yan, Huaguang},
  journal={IEEE Transactions on Smart Grid}, 
  title={Carbon Emission Flow From Generation to Demand: A Network-Based Model}, 
  year={2015},
  volume={6},
  number={5},
  pages={2386-2394},
 }

@ARTICLE{10433421,
  author={Gao, Hongchao and Jin, Tai and Wang, Guanxiong and Chen, Qixin and Kang, Chongqing and Zhu, Jingkai},
  journal={Journal of Modern Power Systems and Clean Energy}, 
  title={Low-Carbon Dispatching for Virtual Power Plant with Aggregated Distributed Energy Storage Considering Spatiotemporal Distribution of Cleanness Value}, 
  year={2024},
  volume={12},
  number={2},
  pages={346-358},
 }

@ARTICLE{8610327,
  author={Celik, Berk and Suryanarayanan, Siddharth and Roche, Robin and Hansen, Timothy M.},
  journal={IEEE Transactions on Sustainable Energy}, 
  title={Quantifying the Impact of Solar Photovoltaic and Energy Storage Assets on the Performance of a Residential Energy Aggregator}, 
  year={2020},
  volume={11},
  number={1},
  pages={405-414},
}

@misc{MEI2020,
  title={Marginal Greenhouse Gas Emission Factors for Ontario Electricity Generation and Consumption},
  author={Travis Lusney},
  year={2020},
}
